# Performance Evaluation of Multi-hop Relaying over Non-Gaussian PLC Channels

Khaled M. Rabie, *Member, IEEE,* Bamidele Adebisi, *Senior Member, IEEE,* Haris Gacanin, *Senior Member, IEEE,* Galymzhan Nauryzbayev, *Member, IEEE,* and Augustine Ikpehai, *Student Member*

*Abstract*—Relaying over power line communication (PLC) channels can considerably enhance the performance and reliability of PLC systems. This paper is dedicated to study and analyze the energy efficiency of multi-hop cooperative relaying PLC systems. Incremental decode-and-forward (IDF) relying is exploited to reduce the transmit power consumption. The PLC channel is assumed to experience log-normal fading with impulsive noise. The performances of single-hop and conventional DF relaying systems are also analyzed in terms of outage probability and energy efficiency for which analytical expressions are derived. Results show that using more relays can improve the outage probability performance; however, this is achieved at the expense of increased power consumption due to the increased static power of the relays, especially when the total source-to-destination distance is relatively small. Results also demonstrate that the IDF PLC system has better energy efficiency performance compared to the other schemes.

## I. Introduction

THE rising demand for communication services has fueled the rapid development of power line communication (PLC) technology witnessed in recent times. However, the power line channel is naturally unfavourable to communication signals given its intrinsic properties such as frequency-selectivity, high incidence of noise and unpredictable line impedance [1]–[3]. Collectively, these factors may result in low signal-to-noise ratio (SNR) values at the receiver. Notwithstanding, this harsh environment, PLC has continued to gain acceptance in different applications such as smart grid, smart home and other cyber-physical systems [4]–[6]; thanks to advanced signal processing and multi-carrier techniques available today. By using the existing electrical wiring for transmitting data signals, PLC eliminates the need for new physical medium which drastically reduces the cost of deployment. Another key benefit of PLC is that it readily provides an alternative in environments where wireless technologies either fail or are unacceptably poor, e.g., in cellars. According to existing standards (such as IEEE 1901 for HomePlug), the communication between PLC nodes is based on carrier sense multiple access with collision avoidance (CSMA/CA) over time division multiple access (TDMA) technique [7], [8]. While CSMA/CA allows the nodes to contend for access to the power line channel without collision, TDMA ensures contention-free slots to serve applications and services that require deterministic allocation of network resources. This combination improves the power line's suitability for multi-hop topologies [7]. Therefore, by allowing the network nodes to act as potential repeaters, relaying neighbour's messages, nodes mutually benefit from one another and the presence of multiple nodes can be exploited to improve network performance; this is broadly referred to as cooperative relaying[1]. Different forms of cooperative relay techniques have been considered in PLC with varying degrees of performance and constraints. This mainly includes amplify-and-forward (AF) and decode-and-forward (DF) relaying [8], [10], [11]. For example, it was shown in [11] that a dual-hop AF PLC system can remarkably improve the system capacity compared to direct-link (DL) transmission. In addition, the authors of [12] and [13] analyzed the performance of multi-hop AF and DF relaying PLC systems in terms of the end-to-end average bit error rate and ergodic capacity where they showed that PLC systems can be made more reliable by increasing the number of relays. However, increasing the number of PLC modems contributes more to the total power consumption due to the aggregate static power of the modems.

Energy efficiency, similar to wireless communication [14], has recently become a trending topic in PLCs. For instance, the authors of [15] and [16] investigated the power consumption in opportunistic DF relaying PLC systems. In [17], the authors evaluated the energy efficiency performance of a half-duplex DF PLC relaying network and later they extended this to MIMO PLC with DF relaying [18]. In addition, it was shown in [19] that adaptively adjusting the transmission parameters based on intelligent signal detection and resource allocation algorithms can considerably enhance the energy efficiency of PLC systems. Very recently, however, instead of only minimizing the transmit power as in the aforementioned studies, the authors in [20]–[22] have proposed harvesting the impulsive noise energy over PLC channels with various relaying and energy-harvesting protocols where it was shown that considerable gains can be attained over conventional systems. It is worth mentioning that impulsive noise is always

Khaled M. Rabie, Bamidele Adebisi and Augustine Ikpehai are with the School of Electrical Engineering, Manchester Metropolitan University, Manchester, M15 6BH, UK. (e-mails: k.rabie@mmu.ac.uk; b.adebisi@mmu.ac.uk, a.ikpehai@mmu.ac.uk).
Haris Gacanin is a research director with Nokia, Copernicuslaan 50, 2018 Antwerp, Belgium. (e-mail: haris.gacanin@nokia.com).
Galymzhan Nauryzbayev is with the faculty of physics and technical sciences, L.N. Gumilyov Eurasian National University, Astana, Kazakhstan (email: nauryzbayevg@gmail.com).

---

[1]The concept of relaying was originally proposed in wireless systems, see e.g., [9].

detected and then eliminated in PLC systems which can be energy inefficient, see e.g., [23]–[25] and the references therein.

To the best of our knowledge, all the existing pieces of work are limited to single-hop or dual-hop relaying and none has evaluated the energy efficiency of multi-hop PLC systems. Unlike previous work, the focus of this paper is to provide detailed mathematical analysis of the outage probability of multi-hop PLC systems with a view to improving their energy efficiency. We also analyze the performance of incremental DF (IDF) relaying in PLC in the presence of impulsive noise. Throughout this paper, the performance of the single-hop PLC system is included to quantify the achievable gains.

The contributions of this paper are as follows. First, closed-form analytical expressions are derived for the outage probability and energy efficiency of a single-hop PLC system. Note that the outage probability indicates decoding failure due to PLC channel fading and noise effects. After that, accurate analytical expressions are formulated for the outage probability and energy efficiency of a multi-hop DF PLC system. The final contribution is that we measure the impact of various parameters on the system performance; that includes the number of relays, impulsive noise probability, distance, static power and various outage probability requirements. The results reveal that, for a given source-to-destination distance, increasing the number of relays can remarkably reduce the outage probability at the expense of increased power consumption due to the increased static power of the relays. It is also shown that the IDF PLC relaying system can considerably improve the energy efficiency of multi-hop PLC systems and this improvement becomes more pronounced at relatively small distances. In addition, it is found that as the impulsive noise probability or the relay static power increases, the energy efficiency of multi-hop PLC systems degrades.

The rest of this paper is organized as follows. In Section II, the system model is described. Section III provides detailed analysis of the outage probability and energy efficiency of the single-hop, multi-hop and IDF relaying schemes over the log-normal fading PLC channel contaminated with impulsive noise. Numerical examples and simulation results are presented and discussed in Section IV, illustrating the impact of various system parameters on the outage probability and energy efficiency. Finally, the main conclusions are presented in Section V.

## II. SYSTEM MODEL

The multi-hop relaying system investigated in this study is shown in Fig. 1a. This system consists of a source modem (S) and a destination modem (D), between which there are $M$ intermediate relays. The $m^{th}$ relay is denoted as $R_m$ where $m \in [1, M]$ and the end-to-end communication is accomplished via the $M$ relays. The PLC channel is assumed to have log-normal distribution, [26], and the noise is modeled using the well-known two-term Gaussian-Bernoulli model consisting of both the background and impulsive noise components [27]. As mentioned in the introduction, the DL approach, i.e. direct source-to-destination transmission without relaying, is also studied here, a block diagram of which is illustrated in Fig. 1b.

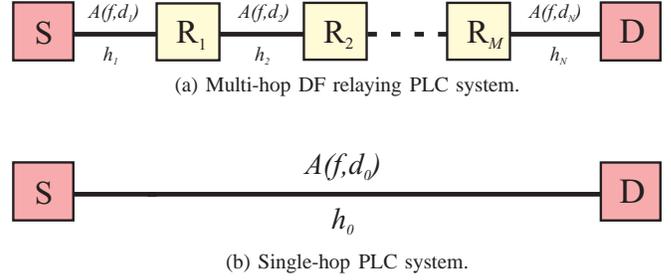

(a) Multi-hop DF relaying PLC system.

(b) Single-hop PLC system.

Figure 1: Basic diagrams of the systems studied in this work.

As a measure of performance, energy efficiency can be approached in a variety of ways. While in the traditional sense it refers to the number of transmitted bits per unit energy, it can also be measured as the transmit energy per bit [28], [29]. Hence, they represent two approaches for solving the same problem. While the goal of the former is to maximize the number of transmitted bit per unit energy, the latter aims to minimize transmit energy per bit; the latter is adopted in this paper. Because our power consumption profile takes into account not only the dynamic power but also the static power of the PLC modems consumed by the circuitry [17], [30], the energy per bit can be determined as

$$E_b = \frac{P_t + P_{stc}^{Tx} + P_{stc}^{Rx}}{R_b}, \quad (1)$$

where $E_b$ is the energy/bit, $P_t$ is the source transmit power for a given outage probability, $R_b = \xi B$ is the data rate in bits/s, $\xi$ is the spectral efficiency and $B$ is the system bandwidth, in Hz, which is assumed to be 30 MHz in all our evaluations, $P_{stc}^{Tx}$ and $P_{stc}^{Rx}$ are the static powers of the transmitting and receiving PLC modems, respectively[2]. It should be highlighted that in order to minimize the energy consumption, the transmit power must be minimized since the static powers are circuitry-specific. Below, we briefly discuss the channel and noise models deployed in this work.

### A. Channel Model

The channel coefficients and the corresponding distances of the multi-hop system are denoted respectively as $h_n$ and $d_n$ as illustrated in Fig. 1a, where $n \in \{0, 1, .., N\}$ and $N$ represents the number of hops, i.e. $N = M + 1$. For the DL system, the channel coefficient and the source-to-destination distance are denoted as $h_0$ and $d_0 = \Sigma_{n=1}^{N} d_n$, respectively. The channels are assumed to be independent and identically distributed following log-normal distribution with a probability density function (PDF)

$$f_Z(z_n) = \frac{\zeta}{\sqrt{2\pi}\sigma_n z_n} \exp\left(-\frac{(10\log_{10}(z_n) - \mu_n)^2}{2\sigma_n^2}\right), \quad (2)$$

---

[2] It is assumed that all the PLC modems have identical power consumption features and therefore $P_{stc}^{Tx}$ and $P_{stc}^{Rx}$ are equal for all modems.

where $z_n = h_n^2$, $n \in \{0, 1, .., N\}$, $\zeta = 10/\ln(10)$ is a scaling constant, $\mu_n$ and $\sigma_n$ (both in decibels) are the mean and the standard deviation of $10\log_{10}(h_n)$, respectively. In addition, the PLC channel suffers from high distance- and frequency-dependent attenuation and losses. This effect is also considered in our analysis and is referred to as $A(f, d)$, where $d$ is the distance and $f$ is the operating frequency.

### B. Noise Model

To accurately characterize the PLC channel impairments, the noise at all modems is assumed to consist of both background noise and impulsive noise. These noise types are modeled using the Gaussian-Bernoulli noise model, [27], in which the background component, $n_w$, is considered complex Gaussian with zero mean and variance $\sigma_w^2$, whereas the impulsive part, $n_i$, is modeled as a Bernoulli-Gaussian random process. Hence, $n = n_w + n_i$, where $n$ is the total noise, $n_i = \mathtt{b}\, g$, $g$ is complex white Gaussian noise with mean zero and $\mathtt{b}$ is the Bernoulli process with $\Pr(\mathtt{b}=1) = p$ with $p$ representing the probability occurrence of impulsive noise. Therefore, the PDF of the total noise can be simply expressed as

$$f_n(n) = \sum_{j=0}^{1} p_j \, \mathcal{CN}\left(n, 0, \sigma_j^2\right), \quad (3)$$

where $p_0 = 1 - p$, $p_1 = p$, $\mathcal{CN}(\cdot)$ denotes the Gaussian PDF, $\sigma_0^2 = \sigma_w^2$, $\sigma_1^2 = \sigma_w^2 + \sigma_i^2$ and $\sigma_i^2$ is the impulsive noise variance. The variances $\sigma_w^2$ and $\sigma_i^2$ define the input signal-to-background noise ratio (SBNR) and the signal-to-impulsive noise ratio (SINR) as follows: $\text{SBNR} = 10\log_{10}\left(1/\sigma_w^2\right)$ and $\text{SINR} = 10\log_{10}\left(1/\sigma_i^2\right)$, respectively. Without loss of generality, we assumed throughout our investigations that the noise statistical characteristics are identical at all PLC modems. It is worthwhile pointing out at this stage that narrow-band noise due to wireless interference is not explicitly considered in this work; for more details on this topic, the reader may refer to [19].

## III. PERFORMANCE ANALYSIS

This section analyzes the outage probability and energy efficiency performance. For better understanding, we first investigate the performance of a single-hop PLC system.

### A. Single-Hop PLC System

In this system, only two modems are engaged in the end-to-end communication, namely, source and destination. Therefore, using (1), the energy consumed per bit in the single-hop system can be expressed as

$$E_{b,SH} = \frac{P_{t,SH} + P_{stc}^{Tx} + P_{stc}^{Rx}}{R_b}, \quad (4)$$

where $P_{t,SH}$ is the source transmit power. To determine $P_{t,SH}$ for a given outage probability, we first need to derive the outage probability for this system as follows. The received signal at the destination, $y_D$, can be written as

$$y_D = P_{t,SH} A(f, d_0) h_0^2 + n_w + n_i. \quad (5)$$

Hence, the SBNR at the destination can be given by

$$\gamma = \frac{P_{t,SH} A(f, d_0) h_0^2}{\sigma_w^2}. \quad (6)$$

With this in mind, the outage probability in the presence of impulsive noise can be calculated as [12], [13], [31]

$$\mathcal{O}^{SH} = \Pr\left\{\sum_{j=0}^{1} p_j \log_2(1+\gamma_j) < \xi\right\}, \quad (7)$$

where $\gamma_0 = \gamma$, $\gamma_1 = \gamma/\beta$ and $\beta = 1 + \sigma_i^2/\sigma_w^2$.

To simplify the analysis, we assume here high SNR approximation. Hence, equation (7) can be approximated as

$$\begin{aligned}\mathcal{O}^{SH} &\simeq \Pr\left\{\log_2(\gamma)^{1-p} + \log_2\left(\frac{\gamma}{\beta}\right)^p < \xi\right\} \\ &\simeq \Pr\left\{\gamma < \beta^p 2^\xi\right\},\end{aligned} \quad (8)$$

which is equivalent to

$$\mathcal{O}^{SH} \simeq F_\gamma\left(\beta^p 2^\xi\right), \quad (9)$$

where $F_\gamma(\cdot)$ is the cumulative distribution function (CDF) of $\gamma$.

It is known that the CDF of a log-normally distributed random variable $(X^2)$ can be given by

$$\begin{aligned} F_X(x) &\simeq \frac{1}{2} + \frac{1}{2}\,\text{erf}\left(\frac{\zeta \ln(x) - 2\mu}{\sqrt{8}\sigma}\right) \quad (10)\\ &\simeq 1 - Q\left(\frac{\zeta \ln(x) - 2\mu}{2\sigma}\right), \quad (11) \end{aligned}$$

where $\mu$ is the mean, $\sigma$ is the standard deviation, $\text{erf}(\cdot)$ and $Q(\cdot)$ denote the error function and the Q-function, respectively.

Based on this definition, and since $h_0^2$ is log-normally distributed (hence, $\gamma$ also has log-normal distribution), the outage probability in (9) can be written as

$$\mathcal{O}^{SH} \simeq \frac{1}{2} + \frac{1}{2}\,\text{erf}\left(\frac{\zeta \ln\left(\beta^p 2^\xi\right) - 2\mu_{h_0} - \zeta \ln(\Lambda)}{\sqrt{8}\sigma_{h_0}}\right), \quad (12)$$

where

$$\Lambda = \frac{P_{t,SH} A(f, d_0)}{\sigma_w^2}. \quad (13)$$

Now, for a given outage probability requirement $(\mathcal{O}^*)$, we can show that the optimal transmit power is given by

$$P_{t,SH}^* = \frac{\beta^p 2^\xi \sigma_w^2}{A(f, d_0)} \exp\left(-\frac{\sqrt{8}\sigma_{h_0}\text{erf}^{-1}(2\mathcal{O}^* - 1) + 2\mu_{h_0}}{\zeta}\right) \quad (14)$$

Finally, the energy per bit for the single-hop system can be obtained by substituting (14) into (4).

## B. Multi-Hop PLC System

In this section, we analyze the outage probability and energy efficiency of various multi-hop relaying scenarios. It should be pointed out that relays are assumed to be spaced equally between the source and destination modems; that is $d_n = d/N$ where $n \in \{1, 2, .., N\}$.

*1) Outage Probability and Energy Efficiency when $N = 2$:* For the case when there are two hops in the system, the outage probability is expressed as

$$\mathcal{O}^{MH2} = \mathcal{O}_{SR_1} + \mathcal{O}_{SR_1}^c \mathcal{O}_{R_1 D}, \tag{15}$$

where $\mathcal{O}_{ij}^c$ is the complement of $\mathcal{O}_{ij}$, i.e. $\mathcal{O}_{ij}^c = 1 - \mathcal{O}_{ij}$ and $\mathcal{O}_{SR_1}$, and $\mathcal{O}_{R_1 D}$ are the outage probabilities of the source-to-relay and relay-to-destination links. Following the same procedure as in the previous section, it is easy to show that

$$\mathcal{O}_{SR_1} \simeq 1 - Q\left(\frac{\zeta \ln(\beta^p 2^\xi) - 2\mu_{h_1} - \zeta \ln(\Upsilon_1)}{2\sigma_{h_1}}\right), \tag{16}$$

$$\mathcal{O}_{R_1 D} \simeq 1 - Q\left(\frac{\zeta \ln(\beta^p 2^\xi) - 2\mu_{h_2} - \zeta \ln(\Upsilon_2)}{2\sigma_{h_2}}\right), \tag{17}$$

where

$$\Upsilon_i = \frac{P_{MH\text{-}2}\, A(f, d_i)}{\sigma_w^2}, \tag{18}$$

with $i \in \{1, 2\}$ and $P_{MH\text{-}2}$ being the transmit power of the two-hop relaying system.

Now, replacing $\mathcal{O}^{MH2}$ in (15) with $\mathcal{O}^*$, we can numerically find the optimal value of $P_{MH\text{-}2}$ ($P_{MH\text{-}2}^*$). Substituting $P_{MH\text{-}2}^*$ into (19)−with $N = 2$−and then into (20) yields the energy consumed per bit for this system.

$$\Gamma_{MH\text{-}N} = \frac{P_{MH\text{-}N}^* + P_{stc}^{Tx} + P_{stc}^{Rx}}{R_b}, \tag{19}$$

$$E_{b,MH\text{-}2} = \mathcal{O}_{SR_1} \Gamma_{MH\text{-}2} + \mathcal{O}_{SR_1}^c 2\Gamma_{MH\text{-}2}. \tag{20}$$

It is worth noting that the first term in (20) indicates the energy consumption when the decoding at the relay is unsuccessful, i.e. lost packets and this energy is wasted. On the other hand, the second term represents the energy usage for successful detection at the relay, i.e. packets are forwarded successfully to the destination.

*2) Outage Probability and Energy Efficiency when $N = 3$:* In this configuration, the overall outage probability can be calculated as follows

$$\mathcal{O}^{MH3} = \mathcal{O}_{SR_1} + \mathcal{O}_{SR_1}^c \left(\mathcal{O}_{R_1 R_2} + \mathcal{O}_{R_1 R_2}^c \mathcal{O}_{R_2 D}\right), \tag{21}$$

where $\mathcal{O}_{SR_1}$, $\mathcal{O}_{R_1 R_2}$ and $\mathcal{O}_{R_2 D}$ represent the outage probabilities of the source-to-relay1, relay1-to-relay2 and relay2-to-destination links, respectively. For the sake of brevity, the derivation of those probabilities are omitted in this paper. These probabilities are given by

$$\mathcal{O}_{SR_1} \simeq 1 - Q\left(\frac{\zeta \ln(\beta^p 2^\xi) - 2\mu_{h_1} - \zeta \ln(\Xi_1)}{2\sigma_{h_1}}\right), \tag{22}$$

$$\mathcal{O}_{R_1 R_2} \simeq 1 - Q\left(\frac{\zeta \ln(\beta^p 2^\xi) - 2\mu_{h_2} - \zeta \ln(\Xi_2)}{2\sigma_{h_2}}\right), \tag{23}$$

$$\mathcal{O}_{R_2 D} \simeq 1 - Q\left(\frac{\zeta \ln(\beta^p 2^\xi) - 2\mu_{h_3} - \zeta \ln(\Xi_3)}{2\sigma_{h_3}}\right), \tag{24}$$

where

$$\Xi_i = \frac{P_{MH\text{-}3}\, A(f, d_i)}{\sigma_w^2}, \tag{25}$$

and $i \in \{1, 2, 3\}$. Replacing $\mathcal{O}^{MH3}$ in (21) with $\mathcal{O}^*$, and using (22)-(25), the optimal value of $P_{MH\text{-}3}$ ($P_{MH\text{-}3}^*$) can be numerically calculated. Then, substituting $P_{MH\text{-}3}^*$ into (19)−with $N = 3$−and then into (26), we can find the energy consumed per bit of this system.

$$E_{b,MH\text{-}3} = \mathcal{O}_{SR_1} \Gamma_{MH\text{-}3} + \mathcal{O}_{SR_1}^c \mathcal{O}_{R_1 R_2} 2\Gamma_{MH\text{-}3} + \mathcal{O}_{SR_1}^c \mathcal{O}_{R_1 R_2}^c 3\Gamma_{MH\text{-}3}. \tag{26}$$

where $\Gamma_{MH\text{-}3}$ is given by (19) when $N = 3$.

*3) Generic Outage Probability and Energy Efficiency with $N$-hops:* The end-to-end outage probability for a network with $N$ hops can be calculated as

$$\mathcal{O}^{MH} = \mathcal{O}_{SR_1} + \left[\sum_{m=1}^{M-1}\left(\mathcal{O}_{R_m R_{m+1}} \times \prod_{i=1}^{m-1} \mathcal{O}_{R_i R_{i+1}}^c\right) \right. \\ \left. + \mathcal{O}_{R_M D} \times \prod_{m=1}^{M-1} \mathcal{O}_{R_m R_{m+1}}^c\right] \times \mathcal{O}_{SR_1}^c, \tag{27}$$

where

$$\mathcal{O}_{SR_1} \simeq 1 - Q\left(\frac{\zeta \ln(\beta^p 2^\xi) - 2\mu_{h_1} - \zeta \ln(\Phi_1)}{2\sigma_{h_1}}\right) \tag{28}$$

$$\mathcal{O}_{R_m R_{m+1}} \simeq 1 - Q\left(\frac{\zeta \ln(\beta^p 2^\xi) - 2\mu_{h_{m+1}} - \zeta \ln(\Phi_{m+1})}{2\sigma_{h_{m+1}}}\right) \tag{29}$$

$$\mathcal{O}_{R_M D} \simeq 1 - Q\left(\frac{\zeta \ln(\beta^p 2^\xi) - 2\mu_{h_{M+1}} - \zeta \ln(\Phi_{M+1})}{2\sigma_{h_{M+1}}}\right) \tag{30}$$

with

$$\Phi_i = \frac{P_{MH\text{-}4}\, A(f, d_i)}{\sigma_w^2}, \tag{31}$$

and $i \in \{1, 2, .., N\}$.

Now, to find the optimal transmit power for a given outage probability ($P_{MH}^*$), we replace $\mathcal{O}^{MH}$ in (27) with $\mathcal{O}^*$ and solve the equation numerically. Substituting the resultant values of $P_{MH}^*$ into (19), and then into (32) gives the energy consumed per bit, where $\Gamma_{MH}$ is given by (19).

$$E_{b,\text{MH}} = \mathcal{O}_{SR_1} \times \Gamma_{\text{MH}} + \mathcal{O}^c_{SR_1}$$
$$\sum_{m=1}^{M-1} \left( (m+1)\mathcal{O}_{R_m R_{m+1}} \Gamma_{\text{MH}} \prod_{i=1}^{m-1} \mathcal{O}^c_{R_i R_{i+1}} \right)$$
$$+ \mathcal{O}^c_{SR_1} \prod_{m=1}^{M-1} \mathcal{O}^c_{R_m R_{m+1}} (M+1)\Gamma_{\text{MH}}. \quad (32)$$

### C. IDF Relaying PLC System

In this system, the relays are only in cooperative mode if the DL does not meet the link quality requirement [32], [33]. To simplify the analysis, we assume that one relay $R_m$ is selected as a cooperative node. Therefore, the overall outage probability of the IDF system is a function of three outage probabilities, given as

$$\mathcal{O}^{IDF} = \mathcal{O}_{SD}\left(\mathcal{O}_{SR_m} + (1 - \mathcal{O}_{SR_m})\mathcal{O}_{R_m D}\right), \quad (33)$$

where $\mathcal{O}_{SD}$ is the outage probability of the source-to-destination link and is given by (12), $\mathcal{O}_{SR_m}$ and $\mathcal{O}_{R_m D}$ are the source-to-relay and relay-to-destination outage probabilities. The relay is assumed to be at the mid-point between the end modems since this usually offers the best performance. Therefore, $\mathcal{O}_{SR_m}$ and $\mathcal{O}_{R_m D}$ can have the follow form

$$\mathcal{O}_x \simeq \frac{1}{2} + \frac{1}{2}\operatorname{erf}\left(\frac{\zeta \ln\left(\beta^p 2^\xi\right) - 2\mu_{h_x} - \zeta \ln(\Psi)}{\sqrt{8}\sigma_{h_x}}\right), \quad (34)$$

where $x \in \{SR_m, R_m D\}$ and

$$\Psi = \frac{P_{IDF}\, A(f, d/2)}{\sigma_w^2}, \quad (35)$$

Using (33), it is easy to show that the optimal transmit power of the IDF system $(P^*_{IDF})$ can be found as the solution of the following equation

$$X + Y + 3Z + XZ + YZ - XY - XYZ = 8\mathcal{O}^* - 3, \quad (36)$$

where

$$X = \operatorname{erf}\left(\frac{\zeta \ln\left(\beta^p 2^\xi\right) - 2\mu_{h_1} - \zeta \ln(\Psi)}{2\sigma_{h_1}}\right), \quad (37)$$

$$Y = \operatorname{erf}\left(\frac{\zeta \ln\left(\beta^p 2^\xi\right) - 2\mu_{h_2} - \zeta \ln(\Psi)}{2\sigma_{h_2}}\right), \quad (38)$$

and

$$Z = \operatorname{erf}\left(\frac{\zeta \ln\left(\beta^p 2^\xi\right) - 2\mu_{h_0} - \zeta \ln(\Lambda)}{2\sigma_{h_0}}\right). \quad (39)$$

Although it is difficult to express $(P^*_{IDF})$ in (36) in closed-form, it is straightforward to find the solution numerically using software tools. Now, to determine the energy consumed per bit, we substitute the optimal transmit power $(P^*_{IDF})$ found from (36) into

$$E_{b,IDF} = \mathcal{O}^c_{SD} \times \frac{P^*_{IDF} + P^{Tx}_{stc} + 2P^{Rx}_{stc}}{R_b}$$
$$+ \mathcal{O}_{SD}\mathcal{O}^c_{SR_m} \times \frac{2P^*_{IDF} + 2P^{Tx}_{stc} + 3P^{Rx}_{stc}}{R_b}$$
$$+ \mathcal{O}_{SD}\mathcal{O}_{SR_m} \times \frac{P^*_{IDF} + P^{Tx}_{stc} + 2P^{Rx}_{stc}}{R_b}. \quad (40)$$

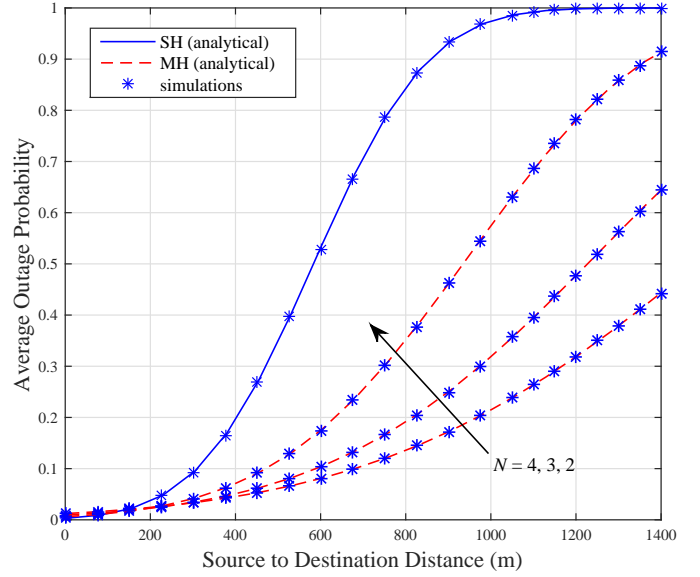

Figure 2: Average outage probability performance of the single-hop and multi-hop relaying PLC systems with various numbers of relays. Relays are spread evenly between the source and destination modems.

## IV. NUMERICAL RESULTS

This section demonstrates some numerical examples of the analytical expressions derived above along with some Monte Carlo simulations. To characterize the distance and frequency-dependent attenuation and losses of the PLC channel, we deploy a common model in which attenuation increases exponentially with distance, given by $A(f,d) = \exp(-\alpha d)$, where $\alpha = a_o + a_1 f^k$ is the attenuation factor, $f$ is the frequency, $k$ is the exponent of the attenuation factor, $a_o$ and $a_1$ are constants determined from measurements and $d$ is the distance. More specifically, we use $a_0 = 9.4 \times 10^{-3}$, $a_1 = 4.2 \times 10^{-7}$, $f = 30$ MHz and $k = 0.7$. In addition, the other system parameters used in this section, unless stated otherwise, are SINR $= -15$ dB, $p = 0.01$, SBNR $= 25$ dB and $\mathcal{O}^* = 10^{-2}$. We also assume that all links have equal variances and means such that $\sigma_n^2 = 2$ dB and $\mu_n = 3$ dB where $n \in \{0, 1, .., N\}$. These values are widely used by many researchers within the PLC community, see e.g., [1], [23], [34] and the references therein.

### A. Average Outage Probability Performance

First, we illustrate in Fig. 2 the analytical and simulated outage probability performance for the multi-hop system with different numbers of relays as a function of the source-to-destination distance; results for a single-hop system are also included. The analytical results for the single-hop approach are obtained from (12) whereas for the multi-hop scheme, the results are obtained using (15), (21) and (27). It should be highlighted that the total source-to-destination distance is kept constant in all scenarios for fair comparison. It is clear from the results in Fig. 2 that the analytical and simulated results of both systems are in perfect match which verifies the correctness of our analysis. It is also obvious that increasing the source-to-destination distance will always degrade performance for all the systems under study and that as we increase the number of relays, for a given distance, the outage probability is enhanced. In addition, it can be seen that this enhancement becomes more obvious as the source and destination modems become more distant and that when the distance is too large, the performance is severely affected irrespective of the number of relays deployed.

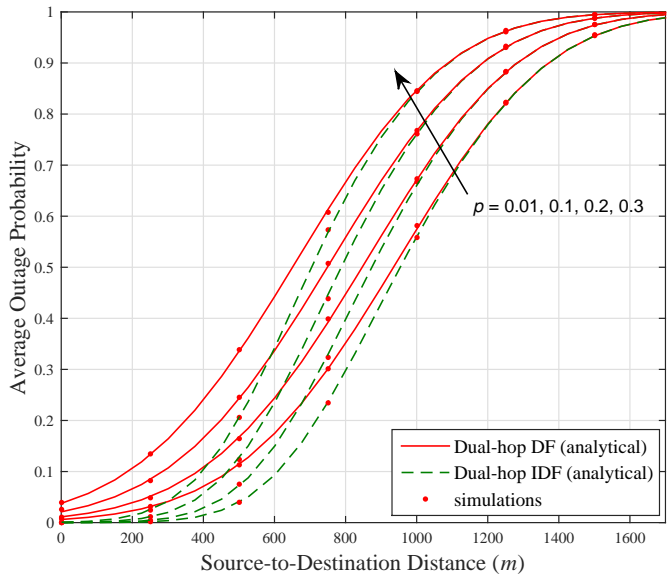

Figure 3: Average outage probability performance of the dual-hop DF and IDF relaying PLC systems with various impulsive noise probabilities.

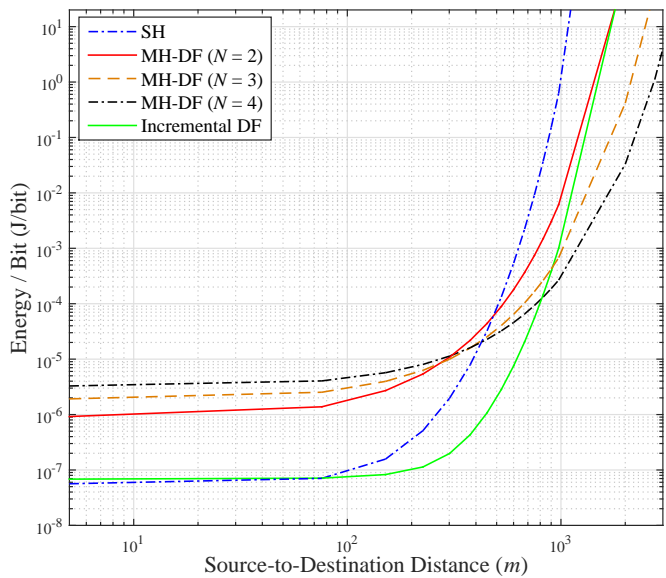

Figure 4: Energy efficiency performance of the multi-hop relaying PLC system with several numbers of relays along with that of IDF and SH systems.

Furthermore, to show the impact of the IDF system on the outage probability in comparison to the conventional DF approach, we plot in Fig. 3 the outage probability of the two systems versus the source-to-destination distance for various impulsive noise probabilities. Again, it is noticeable that the analytical results of the IDF system, obtained from (33), closely match the simulated ones. The other observation one can see from these results is that the IDF system always outperforms the conventional DF scheme for a given pulse probability. More specifically, it is evident that the improvement is more pronounced when the source-to-destination distance is small and as this distance becomes larger, both systems show similar performance. In addition, clearly increasing the noise probability will always worsen the outage performance for the two systems under test.

### B. Energy Efficiency Performance

Although increasing the number of relays will improve the outage probability as shown above, this is obtained at the expense of more energy consumption. In this section, we investigate the energy consumption of the different systems under consideration as shown in Fig. 4. This figure illustrates the energy consumption per bit for the single-hop system, multi-hop system with different relays and the IDF system with respect to the source-to-destination distance. These results are obtained from (14), (20), (26), (32) and (40). A number of important observations can be noticed in this figure. Firstly, it is interesting to see that when the distance is relatively small, in this specific configuration $d_0 = 400$m, the more relays we have the more energy-inefficient the system becomes. In fact, in this region the single-hop approach has the best energy efficiency compared to the other systems. This is because the energy losses due to the static power of the relays outweigh the gains obtained with relaying when the distance is relatively small. On the other hand, however, as the distance becomes larger, the advantage of using relays becomes more pronounced. For instance, it is visible that at $d_0 = 1000$m, the multi-hop system with 3 relays has the best performance whereas the single-hop scheme has the worst energy efficiency. Furthermore, it is noticeable that the IDF has in general better performance at low distance and outperforms all the multi-hop scenarios when the distance is intermediate. However, when the distance is very large, multi-hop systems with 2 and 3 relays outperform the IDF-based scheme.

We now investigate the impact of the static power on the energy efficiency of three systems. Fig. 5 depicts the energy efficiency performance of the single-hop, multi-hop and IDF relaying systems as a function of the static power when $d_0 = 100$m. It is interesting to see the general trend that as the static power of PLC modems increases, all systems become less energy-efficient. In addition, it can be observed that IDF relaying outperforms the multi-hop schemes regardless of the number of relays deployed. However, compared to the single-hop configuration, the IDF system is more energy efficient when the static power is sufficiently low, i.e. lower than 0.9

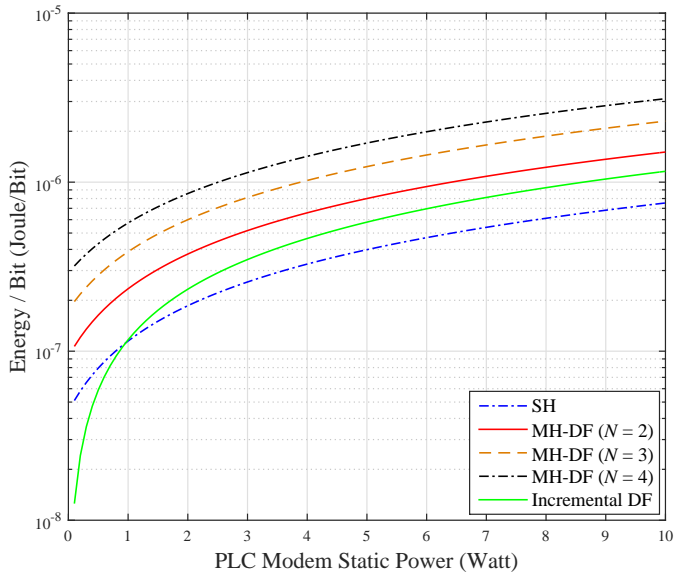

Figure 5: Energy efficiency performance of the single-hop, multi-hop and IDF relaying PLC systems as a function of the relays static power.

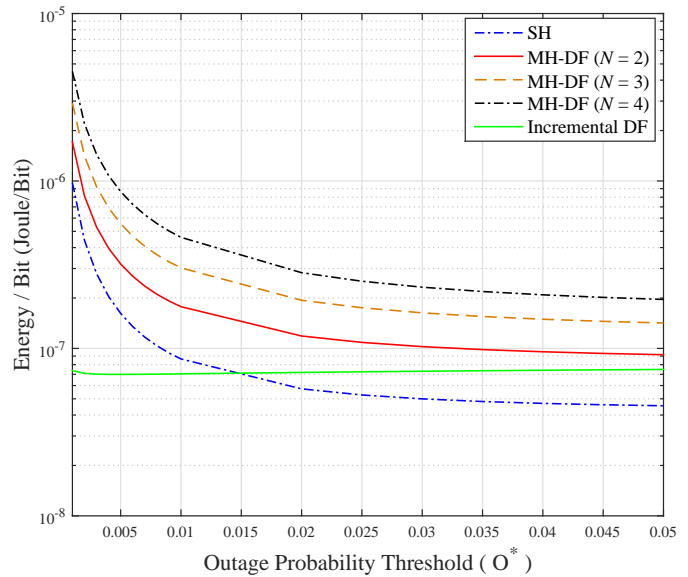

Figure 6: Energy efficiency performance of the single-hop, multi-hop and IDF relaying PLC systems with respect to the outage probability threshold $\mathcal{O}^*$.

Watt in this scenario; whereas when the static power becomes greater than 0.9 Watt, the single-hop approach takes over and becomes more efficient which is justified as discussed previously.

The last set of results in this section is presented in Fig. 6 where the energy efficiency is plotted versus the outage probability threshold for the three systems when $d_0 = 100$m. It is evident from these results that increasing the threshold value leads to higher energy consumption for all the systems, which is intuitive. It should also be pointed out that when the probability threshold is sufficiently high, the change in the energy consumed per bit becomes less significant. The final remark on these results is that IDF relaying is almost independent of the threshold values and that, at low threshold values, the IDF system has best performance relative to other schemes.

## V. Conclusions

This paper studied the performance of multi-hop cooperative relaying PLC systems in terms of the average outage probability and energy efficiency. To improve the energy efficiency in such systems, IDF technique was also analyzed. For comparison's sake, this work included also the performance of single-hop PLC networks. Accurate numerical expressions for the outage probability and energy efficiency for the single-hop, multi-hop and IDF relaying PLC systems were formulated and validated with simulations. Results showed that increasing the number of relays will always improve the outage probability; however, this is achieved at the expense of increased energy consumption since the deployment of more relays implies more static power consumption. It was also presented that the IDF PLC system can provide better energy efficiency compared to the single- and multi-hop systems when the total source-to-destination distance is relatively small.